\documentclass[runningheads]{cl2emult}
\usepackage{subeqnar} % subnumbers individual equations
\usepackage{cropmark} % cropmarks for pages without
\usepackage{eso}      % placeholder for figures
\makeindex            % used for the subject index
\newcommand{\msun}{\mbox{${\rm M}_\odot$}}

\newcommand{\rsun}{\mbox{${\rm R}_\odot$}}

\newcommand{\mbh}{\mbox{$m_{\rm bh}$}}

\newcommand{\rvir}{\mbox{${r_{\rm vir}}$}}

\def\apgt{\ {\raise-.5ex\hbox{$\buildrel>\over\sim$}}\ }
\def\aplt{\ {\raise-.5ex\hbox{$\buildrel<\over\sim$}}\ }
\begin{document}
\title*{Merger rates of black hole binaries:
	prospects for gravitational wave detectors}
\toctitle{The merger rate of black hole binaries}
\titlerunning{The merger rate of black hole binaries}

\author{Simon Portegies Zwart\inst{1} %%\footnote{Hubble Fellow}
\and Stephen L. W. McMillan\inst{2}
}
%
%\authorrunning{Portegies Zwart \& McMillan}
%\slugcomment{Simon Portegies Zwart is Hubble Fellow}
%\footnote{Simon Portegies Zwart is Hubble Fellow}

\institute{Department of Astronomy,
	  Boston University, \\
	  725 Commonwealth Ave., 
	  Boston, MA 02215, USA 
\and 
	Department of Physics,
  	Drexel University, \\
        Philadelphia, PA 19104, USA
}

%---------------------------------------------------------------------
% Removed by Steve -- doesn't work for me, don't know why...
%
%\maketitle              % typesets the title of the contribution
%---------------------------------------------------------------------

\noindent
{\Large Merger rates of black hole binaries: \\
	prospects for gravitational wave detectors} \\

\bigskip

\noindent
{\large Simon Portegies Zwart$^{1, \star}$ and Stephen
L. W. McMillan$^2$} \\

\bigskip
\noindent
$^1$Department of Astronomy, Boston University,  725 Commonwealth Ave., 
	  Boston, MA 02215, USA  \\
$^2$Department of Physics,Drexel University, Philadelphia, PA 19104, USA \\

\bigskip

\noindent
$^\star$ Hubble Fellow \\

\begin{abstract}\noindent
Mergers of black-hole binaries are expected to release large amounts
of energy in the form of gravitational radiation.  However, binary
evolution models predict merger rates too low to be of observational
interest.  In this paper we explore the possibility that black holes
become members of close binaries via dynamical interactions with other
stars in dense stellar systems.  In star clusters, black holes become
the most massive objects within a few tens of millions of years;
dynamical relaxation then causes them to sink to the cluster core,
where they form binaries.  These black-hole binaries become more
tightly bound by superelastic encounters with other cluster members,
and are ultimately ejected from the cluster.  The majority of escaping
black-hole binaries have orbital periods short enough and
eccentricities high enough that the emission of gravitational
waves
causes them to coalesce within a few billion years.  We predict a
black-hole merger rate of $10^{-8}$ to $10^{-7}$ per year per
cubic megaparsec, implying gravity-wave detection rates substantially
greater than the corresponding rates from neutron star mergers.  For
the first generation Laser Interferometer Gravitational-Wave
Observatory (LIGO-I), we expect about one detection during the first
two years of operation.  For its successor LIGO-II, the rate rises to
roughly one detection per day.
There is about an order of magnitude uncertainty in these numbers.

\end{abstract}

\section{Introduction}

Globular clusters contain about one hundred times more low-mass X-ray
binaries (LMXBs) per unit mass than does the Galaxy as a whole---the
Galaxy, with a mass of $2 \times 10^{11}$\,\msun, contains about 100
LMXBs, whereas the Galactic globular cluster population, with a total
mass of just $2 \times 10^{8}$\,\msun, contains at least 10.  All
known cluster LMXBs have neutron stars as primaries.

One might seek an explanation for this discrepancy in LMXB numbers in
the obvious population differences between globular clusters and the
Galactic disc.  The disc contains a mixture of stellar populations,
with broad ranges in age and metallicity, while all stars in a given
globular cluster have essentially the same age and initial
composition.  Conceivably, a globular cluster might experience a
characteristic ``LMXB-rich'' epoch as its component stars evolved.
This hypothesis, however, is not widely accepted.

A more likely explanation for the excess of LMXBs in globular clusters
lies in the radically different dynamics of cluster stars compared to
stars in the Galactic disc.  The mean stellar density in the disc is
about 0.1 star per cubic parsec, with relatively little variation from
place to place.  Globular clusters, on the other hand, exhibit a huge
spread in densities, ranging from values close to the density in the
disc near the cluster tidal radius, to tens of millions of stars per
cubic parsec in the densest cluster cores.  These density differences
may be responsible for the higher birthrate of LMXBs in globular
clusters relative to the Galactic disc---dynamical interactions favor
the formation of LMXBs.

For a cluster age of $\sim$10 Gyr, neutron stars are more than twice
as massive as other cluster members.  Dynamical friction causes them
to sink to the center of the cluster potential well, where stellar
densities are higher and encounters are much more common.  Once in the
core, close encounters with other stars may lead to two-body tidal
capture (Fabian et al.\, 1975) or to three-body exchange interactions
(Phinney \& Sigurdsson 1991). In either case, the neutron star gains a
low-mass companion, which later evolves to become the donor in an
LMXB.  High kick velocities imparted to newborn neutron stars cause
the majority to be ejected from their parent clusters upon formation
(Davies \& Hansen 1998).  Only about 20\% of neutron stars are
retained by globular clusters, yet cluster LMXBs still greatly
outnumber the population in the Galactic disc.  Mass segregation and
tidal capture or exchange are evidently very efficient processes.

Given this reasoning, it is all the more striking that no black-hole
X-ray binaries are observed in globular clusters.  Black holes do not
receive a kick upon formation in a supernova (White \& van Paradijs
1996), so hardly any escape promptly.  Black holes are also
considerably more massive than neutron stars, causing them to sink in
the cluster core even more rapidly.  In equipartition, the black
holes' velocity dispersion is $v \propto m^{-1/2}$.  Thus, the cross
section for a dynamical interaction, which is dominated by
gravitational focusing, is
\begin{equation}
	\sigma \ \propto\  {m \over \sqrt{v}} \ \propto\  m^{5/4}.
\end{equation} 
Hence, for a black hole mass of 10\,\msun, we would naively expect
that globular clusters should contain almost an order of magnitude
more LMXBs with black holes than with neutron stars.  However, none
are found.  The explanation for this discrepancy is as follows.

\section{Black hole formation} 

The initial mass function of globular clusters is well described by a
Scalo (1986) distribution, with lower and upper limits of 0.1\,\msun\,
and 100\,\msun.  This IMF has a mean mass $\langle m \rangle \sim
0.5$\,\msun\, and leads to the formation of about $5\times10^{-4}$
black holes per star.  A $10^6$\,{\msun}\, star cluster thus produces
about 1000 black holes.  Black holes resulting from stellar evolution
are generally quite massive objects: known black hole masses range
from 6 to 10 \msun.  For clarity we adopt a black hole mass of $m_{\rm
bh} = 10$\,\msun; the precise value is not crucial to our discussion,
so long as it significantly exceeds $\langle m \rangle$.

As with neutron stars, dynamical friction causes the black holes to
sink to the cluster core.  The mass segregation time scale is
$\sim\langle m \rangle/m_{\rm bh}$ half-mass relaxation times, or
about $10^8$ yr for $\langle m \rangle = 0.5$\,{\msun} and a cluster
relaxation time of $10^9$ years (see Kulkarni et al.~1992 and
Sigurdsson \& Hernquist 1992 for details).  As mass segregation
proceeds and the cluster core contracts, binaries are formed,
providing the energy needed to support the core against further
gravothermal collapse (Heggie 1975).  The black holes will
preferentially form binaries with one another, both because it is
energetically favorable for them to do so, and because of the
generally larger black-hole interaction cross sections.  Subsequently,
the black-hole binaries evolve via dynamical encounters with other
cluster components.  On average, each encounter between a black-hole
binary and a single black hole hardens the binary (increases its
binding energy) by about 20\%.  Two-thirds of the energy released goes
into binary recoil, the rest into recoil of the other black hole
involved in the interaction.  The hardening process continues until
the recoil velocity exceeds the clusters' escape speed and the
black-hole binary is ejected from the cluster.

A binary can release enough energy to eject itself from the cluster
once its binding energy exceeds $\sim1000$ times the mean kinetic
energy of cluster stars.  By this time the binary has typically
experienced some 40--50 hard encounters.  The recoil energetics imply
that, on average, a black-hole binary is ejected after its {\em
previous} encounter has already ejected a single black hole.  Thus,
for each ejected black-hole binary one expects two single black holes
to be ejected.  There are two possible dynamical scenarios for binary
ejection: (1) there will be at most one or two black-hole binaries in
the core at any given time, and a new binary can form only after these
are ejected; or (2) the core is able to support a large population of
black-hole binaries.  In the former case, the ejection process takes
considerably longer, as the binaries are ejected sequentially.  In the
latter, the binaries may be ejected more or less simultaneously.  Our
simulations are not sufficiently detailed to discriminate between
these alternatives.

In order to eject a black-hole binary following an encounter with a
low-mass cluster member, the binding energy of the black-hole binary
must exceed $\sim4 \times10^4\,kT$.  However, by this time, the
black-hole binary has shrunk to such a small orbital separation that
it likely merges due to emission of gravitational wave radiation
before another encounter takes place.  On the other hand, the
black-hole binary easily ejects low mass stars.  The black hole binary
starts to eject low mass stars as soon as its binding energy exceeds
$\sim 25$\,kT.  At least 20 low mass stars are ejected for each single
black hole.

\section{Characteristics of ejected binaries}

The energy of an ejected binary and its orbital separation are coupled
to the dynamical characteristics of the star cluster.  For a cluster
in virial equilibrium, we have
\begin{equation}
	kT = \frac{2E_{\rm kin}}{3N}
	   = \frac{-E_{\rm pot}}{3N}
	   = {G M^2 \over 6 N \rvir}\,,
\end{equation}
where $M$ and $N$ are the total cluster mass and number of stars,
respectively, and $\rvir$ is the virial radius.  A black-hole binary
with semi-major axis $a$ has
\begin{equation}
	E_b =  {G m_{\rm bh}^2 \over 2 a},
\end{equation}
and therefore 
\begin{equation}
	{E_b \over kT} = 3 N \left( {m_{\rm bh} \over M} \right)^2 
                             {\rvir \over a}.
\label{Eq:Ebhbh}\end{equation}
We can thus compute the properties of black-hole binaries produced by
globular clusters of given masses and virial radii.  These cluster
parameters are assumed to be distributed as independent Gaussians with
means and dispersions of $\log_{10}M = 5.5\pm0.5$ and $\log r_{\rm
vir} = 0.5\pm0.3$, respectively (Djorgovski \& Meylan
1994)\nocite{DM94}.  A recent parameter-space survey of cluster
initial conditions (Takahashi \& Portegies Zwart 2000) \nocite{TPZ20}
finds that typical globular clusters which have survived for a Hubble
time have lost $\apgt 60$\% of their initial mass and have expanded by
about a factor of three.  We correct for this by changing the adopted
distributions to $\log_{10}M = 6.0\pm0.5$ and $\log r_{\rm vir} =
0\pm0.3$.

\section{Production of gravitational radiation}
An approximate formula for the merger time of two stars due to the
emission of gravitational waves is given by 
Peters \& Mathews (1963): \nocite{PM63}
\begin{equation}
 t_{\rm mrg} \approx 150\, {\rm Myr}\; 
	\left( {\msun \over \mbh} \right)^{3}
	\left( {a \over \rsun} \right)^4 (1-e^2)^{7/2} \;.
\label{Eq:tmrg}\end{equation}
Here $e$ is the orbital eccentricity of the black hole binary.  About
90\% of the black-hole binaries formed in the cores of star clusters
merge within a Hubble time due to gravitational radiation.  This
fraction is based on the assumption that the binary binding energies
are distributed flat in $\log E_b$ between $1000\,kT$ and $10000\,kT$,
that the eccentricities are thermal, independent of $E_b$ (these
assumptions are supported by detailed $N$-body simulations of smaller
systems), and that the universe is 15\,Gyr old (Jha et al.\
1999).\nocite{Jha99} The specific contribution to the total merger
rate of black-hole binaries from globular clusters is then about 0.04
per star cluster per million years.

\subsection{Merger rate in the local universe}

We estimate the number density of globular clusters in the universe to
be 
\begin{equation}
	\phi_{GC}  \approx 8.4\,h^3\; {\rm Mpc^{-3}},
\end{equation}
where $h = H_0/100~{\rm km\,s}^{-1}{\rm\,Mpc}^{-1}$.
Combining the specific number density of globular clusters with their
contribution to the black hole merger rate results in a total rate
density of black hole mergers in the universe of
\begin{equation}
	{\cal R}_{GC} \approx 3.2\times 10^{-7} h^3\; 
	                {\rm yr}^{-1}\,{\rm Mpc}^{-3}.
\end{equation}
We note that this figure is larger than the current best estimates of
the neutron-star merger rate ${\cal R} \sim 2 \times 10^{-7}\; h^3\,
{\rm yr}^{-1}\,{\rm Mpc}^{-3}$ (Narayan et al. 1991; Phinney 1991;
Portegies Zwart \& Spreeuw 1996)
\nocite{Nar91}\nocite{Phi91}\nocite{PZS96}.

\subsection{LIGO observations}

The current best estimate of the maximum distance within which LIGO-I
can detect an inspiral event is
\begin{equation}  
	R_{\rm eff} \approx 18\,{\rm Mpc}\ 
			\left(\frac{M_{\rm chirp}}{\msun}\right)^{5/6}
\end{equation}
(K.~Thorne, private communication).  Here, the ``chirp'' mass for a
binary with component masses $m_1$ and $m_2$ is $M_{\rm chirp} = (m_1
m_2)^{3/5} / (m_1+m_2)^{1/5}$.  For neutron star inspiral, $m_1 = m_2
= 1.4\,\msun$, so $M_{\rm chirp} \approx 1.22\,\msun$, $R_{\rm eff}
\approx 21$ Mpc.  For black-hole binaries with $m_1 = m_2 = \mbh =
10\,\msun$, we find $M_{\rm chirp} \approx 8.71\,\msun$, $R_{\rm eff}
\approx 109$ Mpc, and a LIGO-I detection rate of about 1.7\,$h^3$ per
year.  For $h \sim 0.65$ (Jha 1999),\nocite{Jha99} this results in
about one detection event every two years.  LIGO-II should become
operational by 2007, and is expected to have $R_{\rm eff}$ about ten
times greater than LIGO-I, resulting in a detection rate 1000 times
higher---roughly one event per day.

\section{Discussion}

Black-hole binaries ejected from galactic nuclei, the most massive
globular clusters (masses $\apgt 10^6\,\msun$), and globular clusters
which experience core collapse soon after formation tend to be very
tightly bound, have high eccentricities, and merge within a few
million years of ejection.  These mergers therefore trace the
formation of dense stellar systems with a delay of a few Gyr (the
typical time required to form and eject binaries), making these
systems unlikely candidates for LIGO detections, as the majority
merged long ago.  This effect may reduce the current merger rate by an
order of magnitude, but more sensitive future gravitational wave
detectors may be able to see some of these early universe events.  In
fact, we estimate that the most massive globular clusters contribute
about 90\% of the total black hole merger rate.  While their
black-hole binaries merge promptly upon ejection, the longer
relaxation times of these clusters mean that binaries tend to be
ejected much later than in lower mass systems.  Consequently, we have
retained these binaries in our merger rate estimate.

We have assumed that the mass of a stellar black hole is 10\,\msun.
Increasing this mass to 18\,{\msun}\, decreases the expected merger
rate by about 50\%---higher mass black holes tend to have wider
orbits.  However, the larger chirp mass increases the signal to noise,
and the distance to which such a merger can be observed increases by
about 60\%.  The detection rate on Earth therefore increases by about
a factor of three.  For 6\,{\msun} black holes, the detection rate
decreases by a similar factor.  For black-hole binaries with component
masses $\apgt 12$\,\msun, the first generation of detectors will be
more sensitive to the merger itself than to the inspiral phase that
precedes it (Flanagan \& Hughes 1998)\nocite{FH98}.  Since the
strongest signal is expected from black-hole binaries with high-mass
components, it is critically important to improve our understanding of
the merger waveform.  Even for lower-mass black holes (with
$m_{bh}\apgt 10\,\msun$), the inspiral signal comes from an epoch when
the holes are so close together that the post-Newtonian expansions
used to calculate the wave forms are unreliable.  The wave forms of
this ``intermediate binary black hole regime'' (Brady et
al. 1998)\nocite{BCT98} are only now beginning to be explored.
Finally we stress that the black-hole binaries are highly eccentric,
which affects their gravitational wave signals and also influences
their detectability.

\bigskip\noindent{\bf Acknowledgments} We thank Piet Hut, Jun Makino
and Kip Thorne for insightful comments on this work.  This work was
supported by NASA through Hubble Fellowship grant HF-01112.01-98A
awarded (to SPZ) by the Space Telescope Science Institute, which is
operated by the Association of Universities for Research in Astronomy,
Inc., for NASA under contract NAS\, 5-26555, and by ATP grant
NAG5-6964 (to SLWM).  SPZ is grateful to Drexel University and Tokyo
University for their hospitality and for the use of their GRAPE
systems. Part of the calculations are performed on the SGI/Cray
Origin2000 supercomputer at Boston University.

%INDEX%%%%%%%%%%%%%%%%%%%%%%%%%%%%%%%%%%%%%%%%%%%%%%%%%%%%%%%%%%%%%%%
\clearpage
\addcontentsline{toc}{section}{Index}
\flushbottom
%\printindex
%%%%%%%%%%%%%%%%%%%%%%%%%%%%%%%%%%%%%%%%%%%%%%%%%%%%%%%%%%%%%%%%%%%%%

\end{document}